# Harvesting Light with Transformation Optics

Yu Luo*, Rongkuo Zhao, A. I. Fernandez-Dominguez, Stefan A. Maier, J. B. Pendry

*The Blackett Laboratory, Department of Physics, Imperial College London, London SW7 2AZ, UK*

* Electronic email of the corresponding author: y.luo09@imperial.ac.uk

**Abstract:** Transformation optics (TO) is a new tool for controlling electromagnetic fields. In the context of metamaterial technology, it provides a direct link between a desired electromagnetic (EM) phenomenon and the material response required for its occurrence. Recently, this powerful framework has been successfully exploited to study surface plasmon assisted phenomena such as light harvesting. Here, we review the general strategy based on TO to design plasmonic devices capable of harvesting light over a broadband spectrum and achieving considerable field confinement and enhancement. The methodology starts with two-dimensional (2D) cases, such as 2D metal edges, crescent-shaped cylinders, nanowire dimers, and rough metal surfaces, and has been well extended to fully-fledged three-dimensional (3D) situations. The largely analytic approach gives physical insights into the processes involved and suggests the way forward to study a wide variety of plasmonic nanostructures.

## 1. Introduction

The ability of harvesting light and efficiently concentrating its energy into a deep subwavelength volume is highly desired for many applications, such as biosensing [1-4], single-molecule detection [5-8], nanolasing [9-12], and enhanced high harmonic generation [13-15]. In free space, a photon's energy spreads out spatially over about a cubic wavelength, while electrical and chemical energy are concentrated within volumes several orders of magnitude smaller. Therefore, an efficient conversion process between them requires gathering light at the micron scale and focusing its energy onto a nanoscale hotspot. Subwavelength control of light demands new optical materials, and efforts have turned to metals such as gold and silver, where the collective excitations of the conduction electrons, the so-called surface plasmons, couple to light, making possible the compression of EM energy into just a few cubic nanometers [16-24]. Critical to this harvesting process is the ability to fully characterize and model the plasmonic properties of metallic nanostructures. With a tight control over the nanostructures in terms of size and shape, light can be effectively localized down to the nanometer length scale and manipulated with unprecedented accuracy [25-29].

TO is a new tool for the design of EM devices [30-32]. It is based on the fact that Maxwell's equations can be written in a form-invariant manner under coordinate transformations, provided that the permittivity and permeability tensors are modified according to the following equation [33],

$$\begin{aligned}\varepsilon'^{i'j'} &= [\det(\Lambda)]^{-1} \Lambda_i^{i'} \Lambda_j^{j'} \varepsilon^{ij} \\ \mu'^{i'j'} &= [\det(\Lambda)]^{-1} \Lambda_i^{i'} \Lambda_j^{j'} \mu^{ij}\end{aligned} \quad (1)$$

where $\varepsilon^{ij}$ ($\mu^{ij}$) and $\varepsilon'^{i'j'}$ ($\mu'^{i'j'}$) are the permittivity (permeability) elements in the original and transformed spaces, respectively. $\Lambda_i^{i'} = \partial x'^{i'} / \partial x^i$ is the Jacobian matrix relating the two coordinate frames. By applying a specific coordinate transformation to the constitutive relations, EM waves in a given coordinate system can be described as if propagating in a different frame. The TO approach exploits this geometric interpretation of Maxwell's equations, which provides a powerful and intuitive tool for the manipulation of EM fields at all length scales. Probably, the paradigmatic example of the power of TO is the invisibility cloak proposed by Pendry and co-workers in 2006 [30]. There followed a huge surge of interest both from scientists and engineers implementing their own versions of the technology, and from the popular press where it has inspired many articles on the subject, raising the profile of EM phenomena to levels not previously seen [34-54]. Unlike the inexact technique of ray optics (e.g. Snell's Law), TO has the advantage of a simple interpretation in terms of moving around electric and magnetic fields, whilst at the same time being exact at the level of Maxwell's equations. This fact makes TO ideally suited to the study of subwavelength optical phenomena, such as those taking place in plasmonic devices [55, 56]. Importantly, these can be treated within the so-called, quasi-static approximation, which is valid for systems much smaller than the incoming wavelength.

The main focus of this review article is to provide a brief overview on the exploitation of TO to the theoretical description of the subwavelength EM fields supported by metallic nanosystems. Moreover, we will also illustrate how TO can be used to study rigorously and design systematically plasmonic nanostructures that are difficult to investigate with traditional methods. Specially, we discuss how to apply TO to the design of plasmonic devices capable of harvest light efficiently over a broadband frequency range, both in the visible and the near infrared regimes.

## 2. General strategy

Since our goal is to achieve broadband light harvesting, we can start from an infinite plasmonic system that has an inherent broadband response to EM excitation, such as a metal-dielectric interface. Although such structure does not have the properties we desire, as it does not concentrate EM energy to nanoscale hot spots, we can use TO to shape it into a geometry suitable for the envisaged applications while preserving the spectral characteristics.

In a 2D scenario, we are able to apply *conformal transformations* to the geometry, which ensures that both the electrostatic potentials and the in-plane components of the material permittivities are preserved under the transformation (The permeabilities do change, but they are not relevant for the study of subwavelength nanopartices). For example, Fig. 1(a) illustrates a dipole source located between two semi-infinite silver slabs. In this system, surface plasmons are excited by the point source and transport EM energy away along the metal surfaces. Although this process is spectrally broadband, the energy is dispersed to infinity instead of being concentrated to a single location. Next, we use an inversion transformation, which maps points at infinity to the origin, and vice versa, with all points in between mapped into a reverse order. This leads to a pair of touching cylinders as the one shown in Fig. 1 (b). The light source is now at infinity, giving rise to a plane wave in the touching cylinders geometry. The surface plasmons are now incoming waves focused at the origin. In other words, the EM energy in the transformed geometry is gathered at the surface of the cylinders and accumulated at the touching point. In the inverted system, distances are compressed and the EM waves are crushed. Thus, the wavelength is reduced and the energy density is increased by the compression factor. In Fig. 1 (c), the electric field strength is plotted around the circumference of the cylinders, showing the enormous enhancement of the field taking place in the vicinity of the touching point.

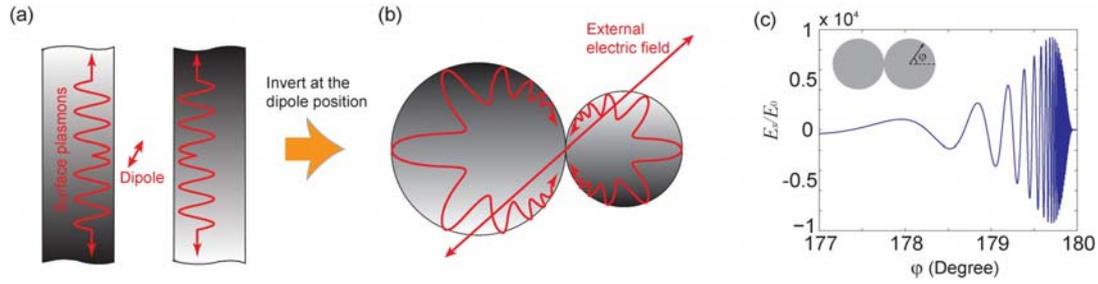

Fig. 1 (a) Two separated semi-infinite slabs of metal illuminated by a dipole source, transporting its energy to infinity. (b) Transformation of the left-hand structure via an inversion about the center of the dipole gives rise to a structure that is finite but inherits the same broadband spectrum. (c) The field enhancement along the circumference of the cylinders in (b).

The efficient harvesting of optical energy can be achieved by a variety of geometries. For each of these structures, which are difficult to treat with traditional EM theories, we can find an appropriate transformation that maps it into a much simpler system, where an analytical description is possible. Following this way, we can inversely engineer the effects of complex light-harvesting nanostructures, and properly design them according to the requirements of the sought-after applications.

## 3. Singular plasmonic structures

Previous theoretical descriptions of the optical properties of metallic nanostructures with sharp geometrical features (or other type of singularities) have shown that these devices can localize light into nanometric hotspots, thereby producing extremely large field enhancements [57, 58]. Sophisticated computer simulations can be used to model these systems [59-66]. However, their numerical treatment, which is very demanding in terms of computational resources, does not provide much physical insight into the phenomena taking place at the geometric singularities. Moreover, each structure configuration requires a new set of simulations, which is extremely time consuming. The TO scheme provides a strategy circumventing this problem, since a single canonical geometry can be related to a large number of transformed devices. For instance, Fig. 2 shows that a periodic metallodielectric system can be linked to a variety of metallic nanostructures, which have different geometric characteristics, but all belong to the same TO family. Normally, the transformed devices have little intrinsic symmetry. For example, the metal crescent-shape cylinder and the sharp edge in Fig. 2 just have a mirror plane along the vertical axis. Hence, computer simulations on these structures often involve a complex representation of the fields. On the contrary, the original planar structures shown in Fig. 2 (a1)-(d1) have translational symmetry along the horizontal direction, enabling us to represent the fields as a simple summation over Bloch wave eigenstates [67, 68]. The solution to this canonical problem can be easily obtained. Then, within the TO framework, we are able to find elegant analytical (or quasi-analytical) solutions to a whole category of nanostructures depicted in Fig.2.

The TO scheme does not only greatly simplify the theoretical investigation of complex nanostructure configurations, but also provides a deep physical insight into the plasmonic phenomena. For instance, it explains why singular plasmonic structures exhibit broadband continuous spectra. As illustrated by Fig. 2, transformations from infinite structures to finite ones give rise to sharp geometrical features. In general, such sharp geometrical features can act as singularities for surface plasmon waves, causing them to slow down as they propagate towards the sharp points. Consequently, light energy builds up around these sharp points and the modal spectrum of the structure become continuous. This effect is shown in Fig. 3, which plots in log scale the broadband continuous absorption cross section of crescent-shaped cylinders (see color lines). For comparison, the narrow spectrum for a single nanowire of the same

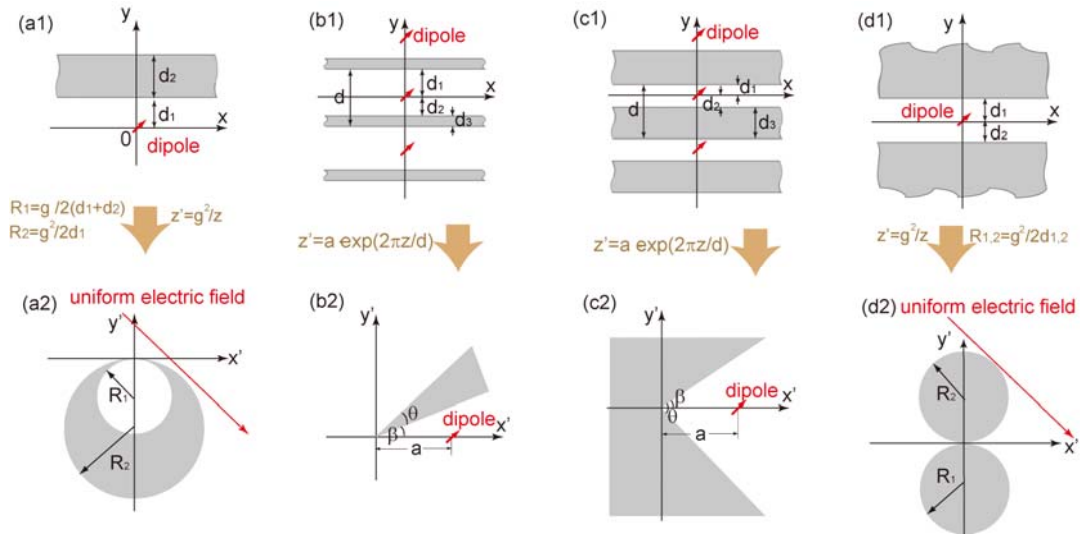

Fig. 2 Schematic of conformal transformations that map canonical metallodielectric system (top panels) to a whole category of singular structures (bottom panels). (a1) A thin metal slab coupling to a 2D line dipole. (a2) A crescent-shaped cylinder illuminated by a uniform electric field. (b1) and (c1) Periodic metallic slabs excited by an array of line dipoles. An exponential transformation converts these two structures into a metallic wedge (b2) and a V-shaped metallic groove (c2) excited by a single line dipole. (d1) Two semi-infinite metal slabs separated by a thin dielectric film and excited by a 2D dipole source. (d2) Two touching metallic nanowires illuminated by a uniform electric field.

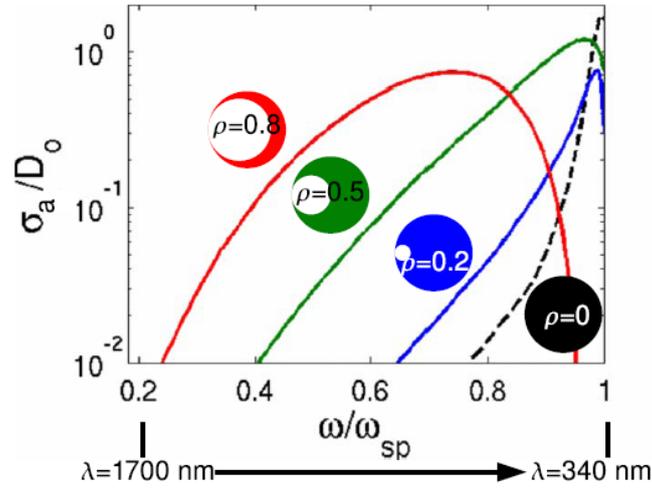

Fig. 3 (solid line) Absorption cross-section normalized to the outer diameter $D_o$, for crescent-shaped cylinders of different geometry. The parameter $\rho$ is defined as the ratio between the inner and outer diameters of the structure. The absorption cross-section for a single nanowire, which shows a narrow peak around the surface plasmon frequency, is plotted in black dashed line. **Figure reprinted with permission: Ref. [69], copyright 2010, APS;**

dimension is rendered in black dashed line. In Fig. 4, we show the field distributions for a series of singular structures under plane wave illumination. In all of them, the surface plasmons excited by the incident fields are efficiently focused, thereby showing extremely large field enhancements in the vicinity of the geometric singularities. It is worth noticing that if the intersection angle at the geometrical singularity is nonzero, the electric field can be divergent even with the inclusion of realistic absorption losses in the metal permittivity [68]. This distinct property is important for Raman scattering spectroscopy. The TO framework enables us to elucidate the optimum shape of the structure that maximizes the Raman signal enhancement. Detailed discussions of this problem can be found in Ref. [70]. Although it is beyond the scope of this review article, the TO framework presented here has been recently extended to describe the emergence of significant nonlocal effects in singular plasmonic geometries. These reflect the occurrence of strong electron-electron interactions that arise in the permittivity of metals structured at the nanoscale. Specifically, the so-called hydrodynamical model was introduced within the TO treatment of nanowire dimers in order to account for the interacting nature of the metal electrons and its impact in the optical response of the device [71, 72].

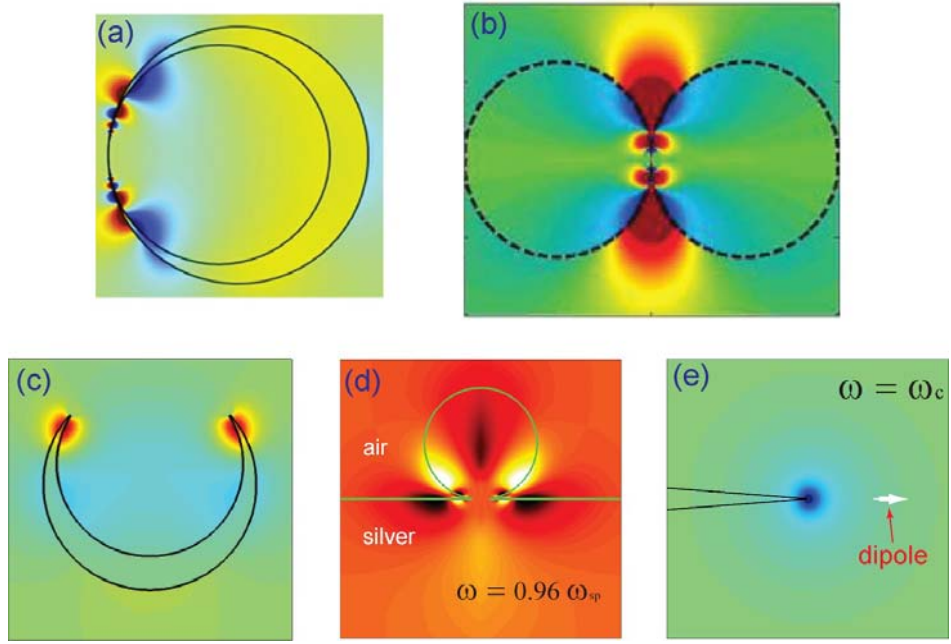

Fig. 4 Electric field distributions for (a) a crescent-shaped cylinder, (b) a pair of touching nanowires, (c) a 2D open-crescent structure, (d) a rough metal surface, (e) a sharp metal wedge. **Figure reprinted with permission: (a), Ref. [69], copyright 2010, APS; (b), Ref. [67], copyright 2010, ACS; (c), Ref. [73], copyright 2012, APS; (d), Ref. [70], copyright 2011, APS; (e), Ref. [68], copyright 2010 ACS.**

## 4. Blunt light harvesting devices

As explained in the previous section, infinite plasmonic systems can be mapped into finite singular structures, which have a continuous spectrum over a broadband frequency. Considering experimental applications, however, these singular structures will suffer from inevitable imperfections due to the nanofabrication process [74-77]. TO enables us to quantitatively examine how the edge rounding at the sharp boundaries of the nanostructures will alter their optical response [73, 78]. Fig. 5 gives two examples of blunt devices (2D blunt crescents and overlapping nanowires) which can be studied within our TO scheme. Since both structures are free from singularities, their 'mother' geometries are no longer infinite, but are truncated in space. As a result, the surface plasmon modes supported by the original structure are quantized at discrete frequencies. The resonance condition in both cases is given by

$$\left(\frac{\varepsilon_m - 1}{\varepsilon_m + 1}\right)^2 \left[\exp\left(\frac{n\pi(2\pi - \theta)}{l_1 + l_2}\right) - \exp\left(\frac{n\pi\theta}{l_1 + l_2}\right)\right]^2 - \left[\exp\left(\frac{2\pi^2 n}{l_1 + l_2}\right) - 1\right]^2 = 0, \quad (2)$$

where $\varepsilon_m$ is the permittivity of the metal; $n$ is an arbitrary integer, denoting the angular momentum of the mode in the transformed frame; and the geometrical parameters $\theta$, $l_1$, and $l_2$ are defined in Fig. 5.

The TO treatment of these systems demonstrated that the resonance frequency and bandwidth of each surface plasmon mode can be controlled by varying the geometry of the blunt edges [73, 78]. Thus, with appropriate design, broadband devices whose absorption properties are robust to edge rounding are possible. Fig. 6 shows that 2D blunt crescents with a tip angle $\theta$ between 30 and 60 degrees exhibit a relatively broadband absorption spectrum, which is nearly independent of the geometrical bluntness [73].

A comprehensive study on structures with asymmetric geometrical bluntness is provided in Ref. [78], which also reports the manifestation of dark modes through symmetry breaking in these devices. It is worth pointing out that the TO based model is computationally much more efficient than numerical simulations. Compared to the commercially available software COMSOL Multiphysics, the theoretical calculations based on TO are faster by a factor of more than $10^3$ when dealing with structures with sharp corners [73, 78]. Importantly, the proposed TO strategy is not restricted to the crescent and overlapping cylinders. By adopting different transformations, it could be used to treat a general class of nanostructures with blunt edges/corners, thereby providing a powerful tool for the design of practical plasmonic devices.

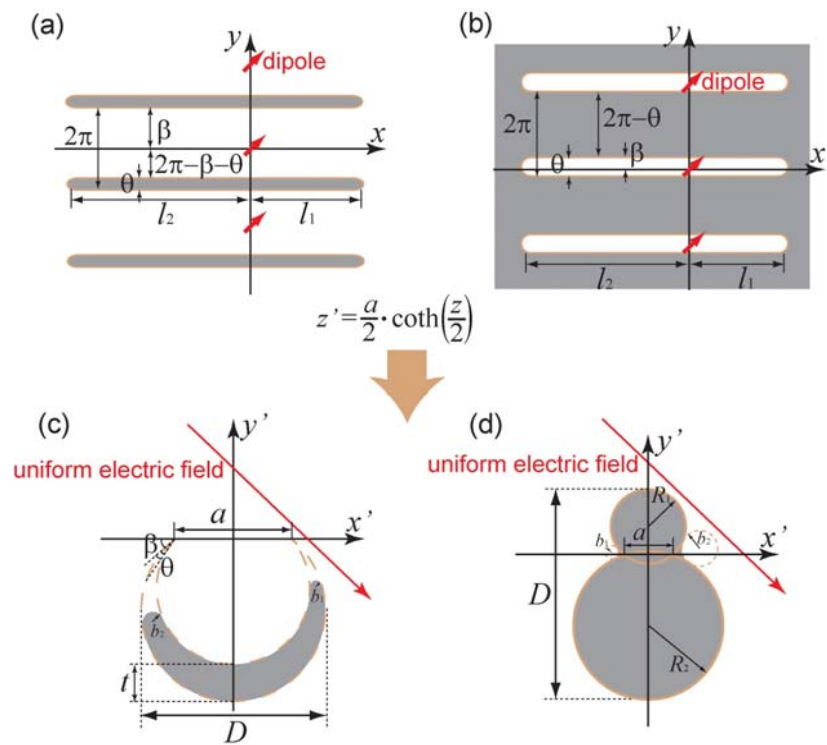

Fig. 5 Panels (a) and (b) show truncated periodic metallodielectric structures. The EM source in both cases is an array of line dipoles (red arrows), aligned along the *y*-axis, with a pitch of $2\pi$. A conformal transformation maps these plasmonic systems into a 2D blunt crescent (c) and a pair of overlapping nanowires (d). The line dipole arrays are transformed into uniform electric fields in both cases. **Figure reprinted with permission: Ref. [78], copyright 2012, ACS.**

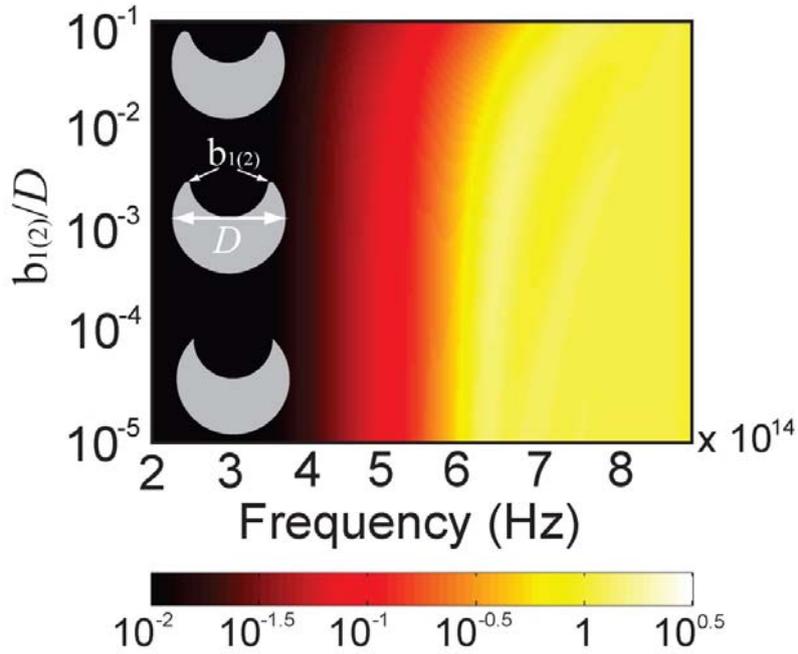

Fig. 6 Absorption cross-section of 2D crescents as a function of the frequency and the geometrical bluntness $b_{1(2)}/D$, where $b_{1(2)}$ is the radius of the corner. **Figure reprinted with permission: Ref. [73], copyright 2012, APS.**

## 5. Interaction between plasmonic nanoparticles

TO can also be applied to study the interaction between plasmonic nanoparticles [79, 80]. In composite systems, the localized surface plasmons supported by individual nanoparticles interact, resulting in a new set of resonances supported by the system as a whole [27-29]. One of the most common approaches used to understand this interaction is the so-called plasmonic hybridization model [81-83]. However, this model still relies on numerical simulations to calculate the exact spectrum of the composite plasmonic structure. In contrast, the TO scheme solves this problem by transforming the interacting nanoparticles into layered structures, on which analytical solutions can be easily found without any fitting parameters [84].

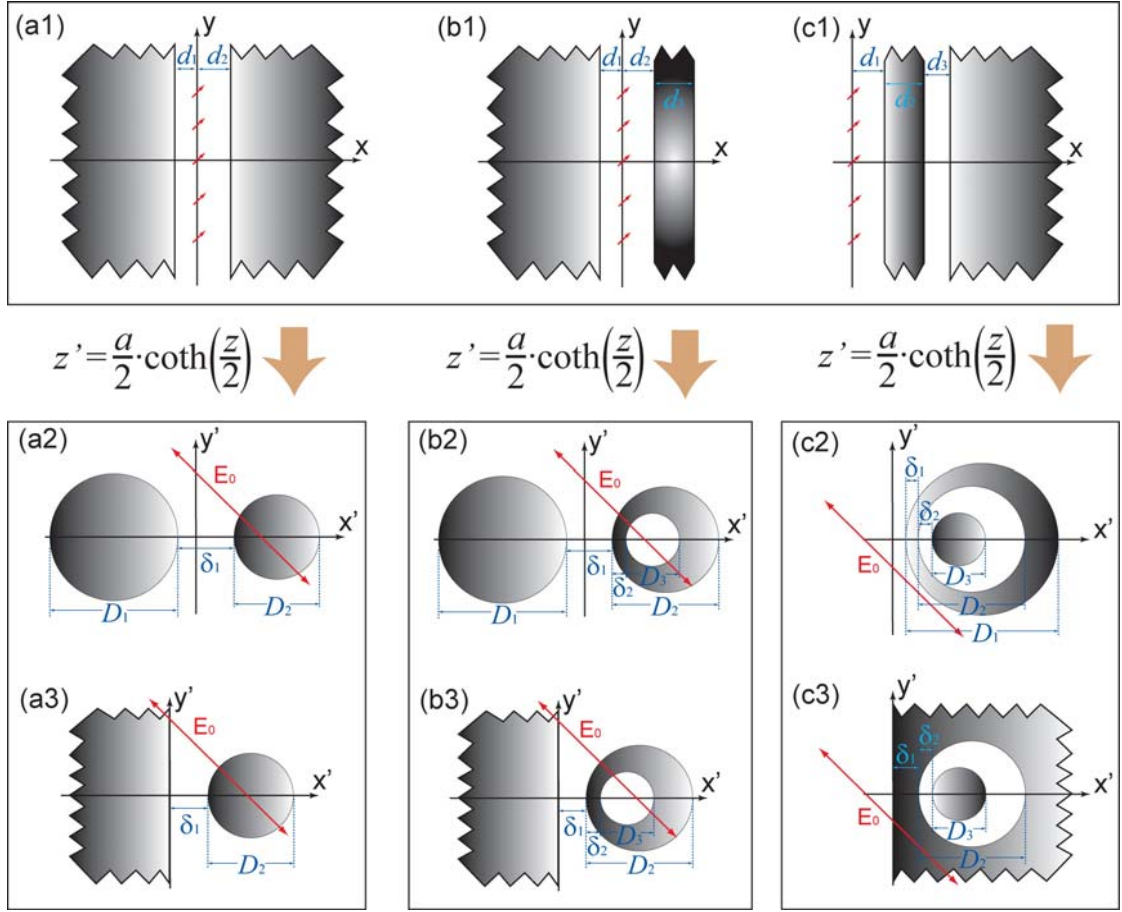

Fig. 7 A conformal transformation $z' = (a/2)\coth(z/2)$ builds bridges between three canonical plasmonic systems (shown in the top three panels) and three types of resonant plasmonic structures. Here the EM source in the initial coordinate frame is an array of line dipoles with a pitch of $2\pi$. Under the conformal mapping, this line dipole array is transformed into a uniform electric field.

Figure 7 depicts typical examples of nanoparticle interactions which can be studied with TO. Since the original plasmonic systems [Fig.6 (a1)-(c1)] are periodic along the *y* axis, the surface plasmon modes supported by these structures are discrete, characterized by linear momenta $k_n = n$ (where *n* is an arbitrary integer). In the transformed frame, these linear momenta correspond to quantized angular momenta of surface plasmon waves that circulate around the nanoparticles. Each angular momentum has a resonance condition, which can be derived from the planar geometry in the original frame [84]. Table 1 summarizes the resonance conditions for the structures shown in Fig. 7 (a2)-(c2), where, *n* denotes the angular momenta associated with different modes, and $\eta$ and $\zeta$ are simple functions of the geometric parameters $\delta$ and *D*.

| Transformed structure | Resonance condition |
|---|---|
| Cylinder pair (Fig. 4.1 (a2)) | $\eta_1^{2n}\eta_2^{2n} = \left(\dfrac{\varepsilon_m - 1}{\varepsilon_m + 1}\right)^2$ |
| Tube/wire dimer (Fig. 4.1 (b2)) | $\left(\dfrac{\varepsilon_m - 1}{\varepsilon_m + 1}\right)^2 \left(\eta_1^{2n}\eta_2^{2n} + \dfrac{\eta_3^{2n}}{\eta_2^{2n}} - 1\right) = \eta_1^{2n}\eta_3^{2n}$ |
| Non-concentric tube/wire cavity (Fig. 4.1 (c2)) | $\left(\dfrac{\varepsilon_m - 1}{\varepsilon_m + 1}\right)^2 \left(\dfrac{\zeta_2^{2n}}{\zeta_1^{2n}} + \dfrac{\zeta_3^{2n}}{\zeta_2^{2n}} - 1\right) = \dfrac{\zeta_3^{2n}}{\zeta_1^{2n}}$ |

Table 1 Resonance conditions for transformed nanostructures depicted in Fig.7 (a2)-(c2).

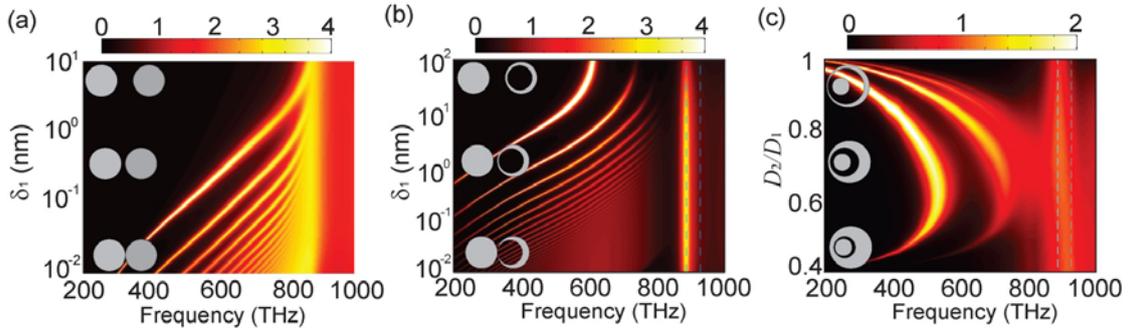

Fig. 8 Evolution of the absorption spectra in terms of the distance between nanoparticles for (a) a pair of separated nanowires, (b) the tube/wire dimer, (c) the tube/wire cavity.

Figure 8 renders the absorption cross section for three nanoparticle configurations as a function of the distance between nanoparticles. The spectra show how the discrete resonance modes supported by these composite structures redshift when the two plasmonic elements approach to each other.

## 6. Extension to 3D geometries

Although the 2D examples discussed in previous sections provide insightful physical understanding to the optical properties of plasmonic nanostructures, they only describe the subset of the spectra for electric fields perpendicular to the cylinder axis (transverse magnetic polarization). Recently, the TO framework has been extended to 3D geometries [85-87], allowing for a complete description of the spectra. In this section, we briefly describe the 3D TO strategy, which exploits the analogy to its 2D counterpart. The example we consider here is a 3D inversion, as shown in Fig. 9. The 3D planar and annulus structures shown in the upper panels have much more symmetry than the structures shown in the bottom panels. Therefore, we can anticipate that the transformation will ease the description of their plasmonic characteristics.

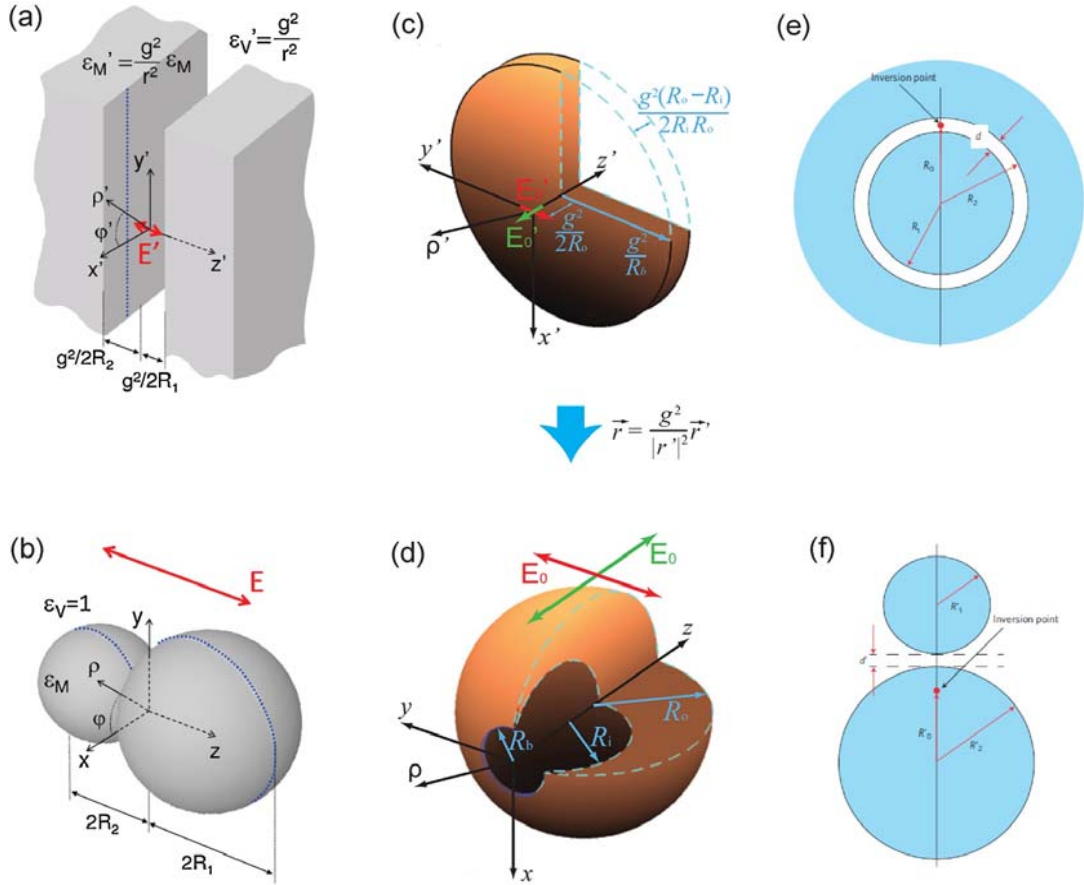

Fig. 9 Panels (a), (c), and (e) depict the starting structures, which have translation/rotational symmetry and whose dielectric functions are space-dependent. Panels (b), (d), and (f) show the transformed geometries, which have little intrinsic symmetry, and are difficult to study using traditional methods. **Figure reprinted with permission: (a) and (b), Ref. [85], copyright 2012, APS; (c) and (d), Ref. [86], copyright 2012, ACS; (e) and (f) Ref. [87], copyright 2013, NPG.**

In contrast to 2D cases, 3D transformations act on both the geometry and material properties of the structure. As the systems in the transformed frame (bottom panels of Fig. 9) comprise homogenous metals immersed in homogeneous dielectrics, the original planar/annulus structures have a space-dependent permittivity of the form

$$\varepsilon'(\mathbf{r}) = g^2 |\mathbf{r}|^{-2} \varepsilon \qquad (3)$$

where $\mathbf{r}$ is the distance to the origin; $g$ is a free parameter of the transformation; $\varepsilon$ denotes the permittivity distribution in the transformed frame. Although the spatial-dependent dielectric function in Eq. 3 complicates the treatment of the 3D problem, a detailed study shows that the quasistatic potential in the original coordinate frame takes the form:

$$\phi(\mathbf{r}) = |\mathbf{r}| V(\mathbf{r}) \quad (4)$$

where $V(\mathbf{r})$ is a solution of Laplace's equation, satisfying $\nabla^2 V(\mathbf{r}) = 0$. Then, by matching the boundary conditions for the simplified geometries in the original frame, we can obtain $\phi(\mathbf{r})$, and hence find (quasi-)analytical solutions for the structures shown in Fig. 9 (b), (d), and (f).

Detailed comprehensive theoretical analysis exploiting TO ideas showed that 3D structures show a better nanofocusing performance than their 2D counterparts [85, 86], yielding larger field enhancements which can reach up to $10^4$ close to the geometrical singularities. This is due to the fact that surface plasmons are focused onto a point in 3D rather than a line, as it happens in 2D. Besides, the 3D solution gives a complete description of the spectra, enabling us to go beyond the description of the spectroscopic properties of the system to other phenomena, such as van der Waals interactions. Detailed discussions of this problem can be found in Ref. [87].

## 7. Experimental realization

Despite the theoretical efforts to reduce the acute geometric features of the TO designs presented above, their experimental realization remains a challenge. On the one hand, the fabrication of these devices requires an enormous precision in order to define accurately the nanostructure geometry. On the other hand, their full optical characterization involves the probing of the local field enhancement with spatial resolution below the nanometer scale. Here, we discuss briefly two works which, going along these lines, found experimental evidence of the TO predictions for the close encounter for two plasmonic nanoparticles.

The left panel of Fig. 10(a) shows the dark-field scattering spectra measured for different incident polarizations (see insets) from a 155 nm diameter gold nanospheres on top of a gold film [88]. The nanoparticles were grown via self-assembly on the substrate film by depositing a thin layer of Au on top of a rough seed layer of $VO_2$. The spectra present three scattering maxima within the wavelength range between 500

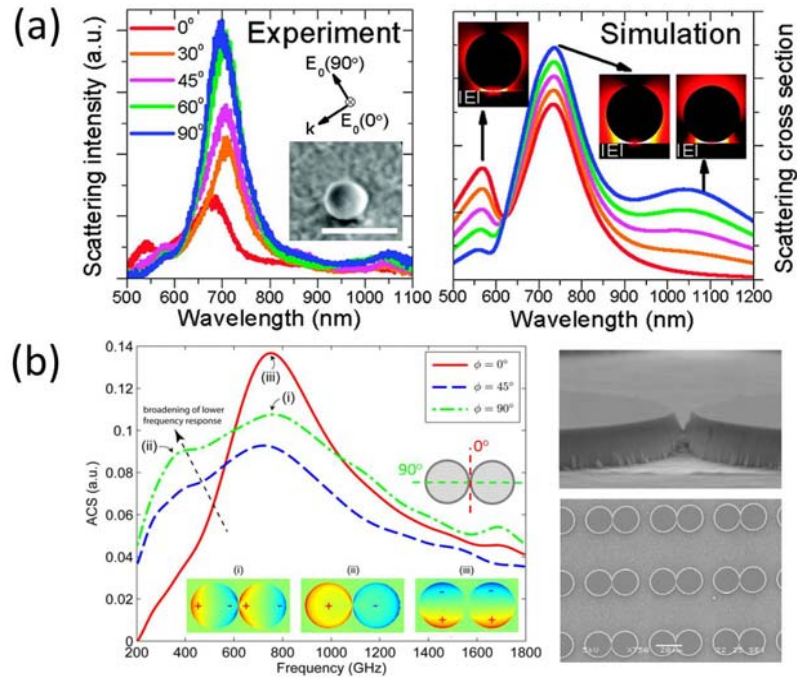

Fig. 10 (a) Measured (left) and calculated (right) dark-field scattering spectra for a 155 nm diameter gold sphere standing on top of a gold film. The various spectra correspond to different polarizations of the incident fields. The left inset shows a scanning electron microscopy (SEM) image of the experimental sample (white bar: 500 nm). The right insets render the field enhancement evaluated at the various scattering maxima. (b) Left: measured THz absorption cross section variation with the incident polarization angle for InSb touching disks of 10 micron radius. The labeled absorption peaks correspond to the charge distribution maps in the inset. Right: lateral and top SEM images of the experimental InSb samples. **Figure reprinted with permission: (a) Ref. [88], copyright 2012, ACS; (b) Ref. [89], copyright 2012, Wiley.**

and 1100 nm. The right panel of Fig. 10(a) plots the scattering cross sections obtained through numerical simulations (COMSOL Multiphysics) for a nanosphere separated by a 2 nm gap from the substrate. Note that both the position and shape of the measured scattering maxima is reproduced by the simulations. The insets render the electric field amplitude evaluated at the three resonances supported by the system, demonstrating the strong field enhancements taking place at the gap of the geometry. Through the analogy with the TO predictions for the 2D analogue for this geometry (a cylinder on top of an infinite substrate) [80], the nature of the plasmonic modes behind these scattering peaks is revealed. The long wavelength resonance, which is efficiently excited by $90^\circ$ illumination, is caused by a dipolar mode oriented normally to the substrate. The intermediate maximum, which is apparent for all incident polarizations, correspond to a hybrid mode with bright dipolar components parallel (located at the nanosphere) and normal (at the gap) to the gold film. Finally, the origin of the short wavelength peak is a high order (quadrupole) mode strongly confined at the gap of the geometry.

A strategy to ease the fabrication of TO designs consists in using semiconductor materials as the plasmonic platform [89]. The dielectric properties of semiconductors, such as intrinsic InSb, can be described through a lossy Drude model with a plasma frequency lying at the THz regime [90, 91]. In contrast, the plasma frequency for noble metals such as gold or silver is located at the visible or ultraviolet range of the EM spectrum. Therefore, by replacing the plasmonic material, the operating wavelength of TO devices can be shifted from the submicron (metals) to the submillimeter (semiconductors) regime. Importantly, this makes the realization of the plasmonic phenomena predicted within the TO framework in much larger structures. However, note that the EM characterization of these systems is not simplified this way due to the lack of efficient THz sources and detectors [92].

The left panel of Fig. 10(b) shows the THz absorption cross section measured from a square array of touching dimers of InSb disks for different incident polarizations. The experimental spectra range between 0.2 and 1.1 THz (note that the InSb plasma frequency is 2.4 THz). The 10 micron radius, 1.3 micron height disks were fabricated on top of a semi-insulating GaAs substrate, and the array pitch was set to 50 microns to avoid the near-field and diffraction coupling between neighboring dimers. The gap region at each dimer was patterned using electron beam lithography, giving rise to a V-shaped (~46$^o$) gap between the disks. See the right insets of Fig. 10(b) for a lateral and top SEM images of the experimental samples.

The experimental cross sections in Fig. 10(b) demonstrate the broadening of the absorption spectrum of the InSb disk dimers as the incident polarization is rotated from 0$^o$ (normal to the dimer axis) to 90$^o$ (parallel to the dimer axis). This observation is in agreement with the continuous broadband response obtained within the TO framework for 2D and 3D touching geometries [67, 85]. Moreover, the TO approach enables us to gain physical insight into the plasmonic modes that govern the optical properties of the structure for different incident polarizations (see the charge distribution maps in the left panel of Fig. 10(b)). Note that, whereas the system supports only one resonance under normal illumination, two different modes emerge when the impinging fields are polarized along the dimer axis. It is the spectral overlapping between these two plasmonic resonances which causes the broadening of the experimental absorption cross section.

## 8. Conclusion

TO provides an elegant tool for the study of plasmonic nanostructures, predicting the key features that allow for strong far to near-field EM energy conversion through a prominent field enhancement and localization. Taking into account geometrical bluntness and nonlocal effects, we have shown that the recently developed TO framework has made a step forward towards practical applications. The improvement of the TO approach by considering these physical constrains allows for predicting analytically the optical properties of a large variety of nanostructures, as well as opening the way to the broadband nanofocusing of light. The plasmonic devices proposed here (such as the crescent and touching spheres) can find potential applications in areas as diverse as fluorescence and photoluminescence, super-resolution imaging, Raman scattering spectroscopy, single-molecule detection, and high-harmonic generation. The analytical modeling of these nanostructures is critical for plasmonic research, as it provides profound understanding and accurate estimation of the optical properties of the geometries of interest, as well as assisting in the actual design of nanoparticles. Moreover, apart from yielding deep physical insight, our methodology makes the accurate and efficient computational modeling of EM phenomena possible, which will be of great value in further studies on plasmonic devices and related systems.

**Acknowledgement**

For our support we thank the following: YL, SAM and JBP the Leverhulme Trust; AIFD and JBP the Gordon and Betty Moore Foundation; RKZ the Royal Comssion for the Exhibition of 1851; JBP the AFOSR; SAM the EPSRC